\documentclass[times,twocolumn,final,authoryear]{elsarticle}

\usepackage{ycviu}
\usepackage{framed,multirow}

\usepackage{amssymb}
\usepackage{latexsym}
\usepackage{graphicx}
\usepackage{amsmath}
\usepackage{amssymb}
\usepackage{comment} 
\usepackage{booktabs} 
\usepackage{graphics}
\usepackage{multirow}
\usepackage{cite}
\usepackage{float}
\usepackage{xcolor}
\usepackage{hyperref}
\newcommand{\norm}[1]{\left\lVert#1\right\rVert}

\usepackage{url}
\usepackage{xcolor}
\definecolor{newcolor}{rgb}{.8,.349,.1}

\journal{Computer Vision and Image Understanding}

\begin{document}

\thispagestyle{empty}

\clearpage

\ifpreprint
  \setcounter{page}{1}
\else
  \setcounter{page}{1}
\fi

\begin{frontmatter}

\title{Wavelet-based network for high dynamic range imaging}
\author[1]{Tianhong \snm{Dai}$^{*,}$}
\author[3]{Wei \snm{Li}$^{*,}$}
\author[4]{Xilei \snm{Cao}}
\author[3]{Jianzhuang \snm{Liu}}
\author[3]{Xu \snm{Jia}}
\author[3]{Ales \snm{Leonardis}}
\author[3]{Youliang \snm{Yan}}
\author[2]{Shanxin \snm{Yuan}\corref{cor1}}
\cortext[cor1]{Corresponding author}

\address[1]{University of Aberdeen, Aberdeen, AB24 3FX, United Kingdom}
\address[2]{Queen Mary University of London, London, E1 4NS, United Kingdom}
\address[3]{Huawei Noah’s Ark Lab, London, SE10 0ER, United Kingdom}
\address[4]{Huawei Technologies Co., Ltd., Beijing, 100085, China}

\received{1 May 2013}
\finalform{10 May 2013}
\accepted{13 May 2013}
\availableonline{15 May 2013}
\communicated{S. Sarkar}

\begin{abstract}
High dynamic range (HDR) imaging from multiple low dynamic range (LDR) images has been suffering from ghosting artifacts caused by scene and objects motion. Existing methods, such as optical flow based and end-to-end deep learning based solutions, are error-prone either in detail restoration or ghosting artifacts removal. Comprehensive empirical evidence shows that ghosting artifacts caused by large foreground motion are mainly low-frequency signals and the details are mainly high-frequency signals. In this work, we propose a novel frequency-guided end-to-end deep neural network (FHDRNet) to conduct HDR fusion in the frequency domain, and Discrete Wavelet Transform (DWT) is used to decompose inputs into different frequency bands. The low-frequency signals are used to avoid specific ghosting artifacts, while the high-frequency signals are used for preserving details. Using a U-Net as the backbone, we propose two novel modules: merging module and frequency-guided upsampling module. The merging module applies the attention mechanism to the low-frequency components to deal with the ghost caused by large foreground motion.  The frequency-guided upsampling module reconstructs details from multiple frequency-specific components with rich details. In addition, a new RAW dataset is created for training and evaluating multi-frame HDR imaging algorithms in the RAW domain. Extensive experiments are conducted on public datasets and our RAW dataset, showing that the proposed FHDRNet achieves state-of-the-art performance.
\end{abstract}


\end{frontmatter}
\def\thefootnote{*}\footnotetext{These authors contributed equally to this work}
\section{Introduction}
\label{sec:introduction}
High dynamic range (HDR) imaging using multiple low dynamic range (LDR) images as inputs is a technique used in computational photography to generate high-quality HDR images. This technique achieves a large range of luminosity by utilizing the information from multiple LDR images. 
A digital camera usually captures an LDR image with only a limited range of luminosity at a time, where there might appear some over-exposed and/or under-exposed regions, degrading the image quality. 
%
%
Cameras embeded in wearable devices usually have small optical sensors and small apertures, which limit the number of electrons to reach each pixel, making them difficult to capture HDR images at a time.
A practical solution for wearable devices is to capture several LDR images with different exposure times and fuse them into a single HDR image. To generate an HDR image, the method should be able to restore the missing information (over-exposed and under-exposed regions) from multiple LDR images, and more importantly, be ghost-free.

\begin{figure}[t]
    \centering
    \includegraphics[width=1.0\columnwidth]{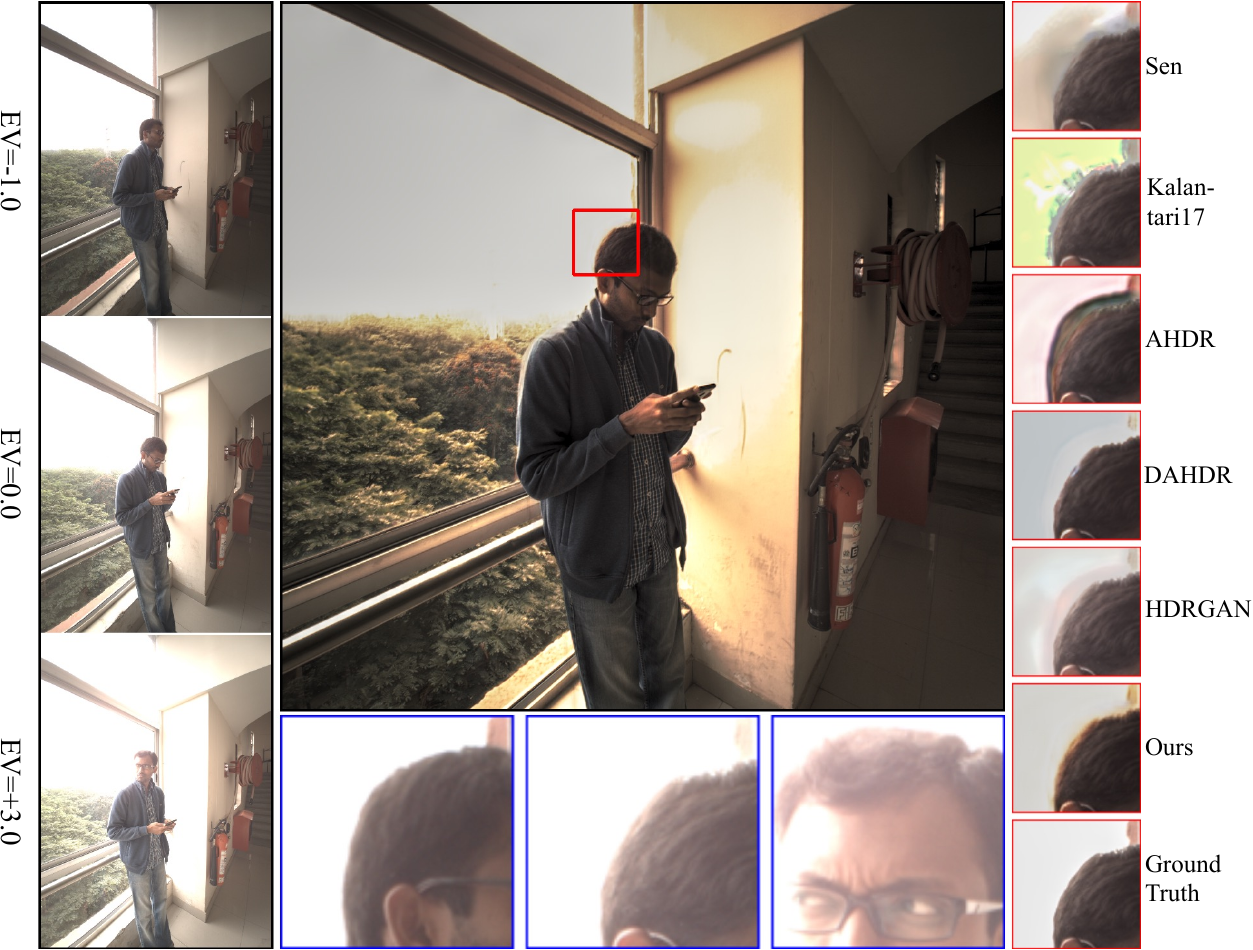}
    \caption{Comparison between our method and other baselines on the Prabhakar dataset~\citep{prabhakar2019fast}. Left: Three LDR images with different exposures (low, medium, and high). Center: Our generated HDR image after tone mapping and cropped LDR patches. Right: Results of six methods and the ground truth.}
    \label{fig:teaser}
     \vspace{-5mm}
\end{figure}

Existing methods~\citep{kalantari2017deep, wu2018deep, yan2020deep, Ram20, yan2019attention, yan2019multi} suffer from different kinds of artifacts, including ghosting, missing details, color degradation, etc. The traditional method by~\citet{debevec1997} can generate a decent high quality HDR image by merging several static LDR images with different exposures, but it might introduce ghosting artifacts when there is motion. Other early works \citep{khan2006ghost, pece2010bitmap, li2014selectively, bogoni2000extending, gallo2015locally} try to deal with motion through detecting and rejecting moving pixels \citep{khan2006ghost, pece2010bitmap, li2014selectively}, or through aligning and merging LDR images~\citep{bogoni2000extending, gallo2015locally}. They can address a small range of motion but they cannot handle moving content effectively.

Recently, deep learning-based methods \citep{kalantari2017deep, wu2018deep, yan2019attention} have been proposed and made great improvements over traditional methods, benefiting from CNN's good representation ability and large amount of training data. These methods either use optical flow to align the inputs, followed by a merging module~\citep{kalantari2017deep}, or formulate the HDR imaging task as an image-to-image translation problem~\citep{wu2018deep, yan2019attention}. Although these methods have made great progress in this area, they still suffer from the ghosting problem (see Figure~\ref{fig:teaser}).
We notice that none of the existing methods tries to exploit the fact that the ghosting artifacts caused by large foreground motion are mainly of low-frequency, while the details are of high-frequency. We argue that it is beneficial to separate these low-frequency and high-frequency signals and deal with them separately. Frequency operation has also been used in a few existing HDR imaging methods~\citep{pouli2014hdr, hasinoff2016burst}, \textit{e.g.}, ~\citet{pouli2014hdr} decomposes HDR frames into different frequency bands, where the most suitable band is selected adaptively to prevent ghosts, ~\citet{hasinoff2016burst} uses pairwise frequency-domain temporal filter operation for a robust and fast alignment.

\begin{figure}[t]
	\centering
	\includegraphics[width=1.0\columnwidth]{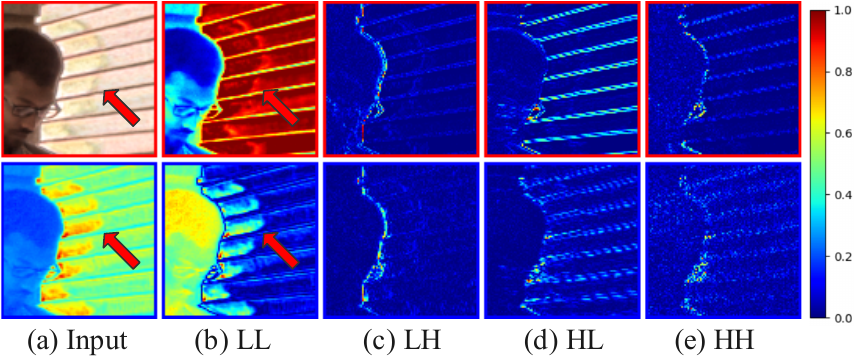}
	\caption{The visualization of frequency sub-bands in the wavelet domain. The first and second row are the results of DWT decomposition of an RGB image and a feature map as the input. (a) Inputs. (b)-(e) Visualization of different frequency sub-bands.}
	\label{fig:motivation}
	\vspace{-5mm}
\end{figure}
In this paper, we choose Discrete Wavelet Transform (DWT) to decompose signals into different frequency bands. Compared with other methods, such as Discrete Fourier Transform (DFT) and Discrete Cosine Transform (DCT), DWT can capture both frequency and spatial information of the images (or feature maps), which helps to preserve detailed texture. In order to verify our hypothesis, we select the output of AHDR~\citep{yan2019attention} in Prabhakar dataset~\citep{prabhakar2019fast}, where a distinct ghosting artifact is presented because of the object motion, as an example to visualize the decomposed signals in each frequency sub-band. After decomposing, the corresponding frequency sub-bands are given in the first row of Figure~\ref{fig:motivation}. It clearly shows that the ghosting artifacts are mainly in the low-frequency sub-band (LL), while the high-frequency sub-bands (LH, HL, HH) include textures in different directions. In order to extend this verification to the feature space, we also investigate the feature map from the last but one layer of AHDR. In the second row of Figure~\ref{fig:motivation}, it presents a similar trend in the feature space where the ghosting artifacts caused by large foreground motion is mainly in the low-frequency sub-band. Thus, it is highly worth exploring frequency-specific processing in both the RGB and deep feature domains for the HDR imaging task.

\begin{figure*}[h!]
    \centering
    \includegraphics[width=0.9\textwidth]{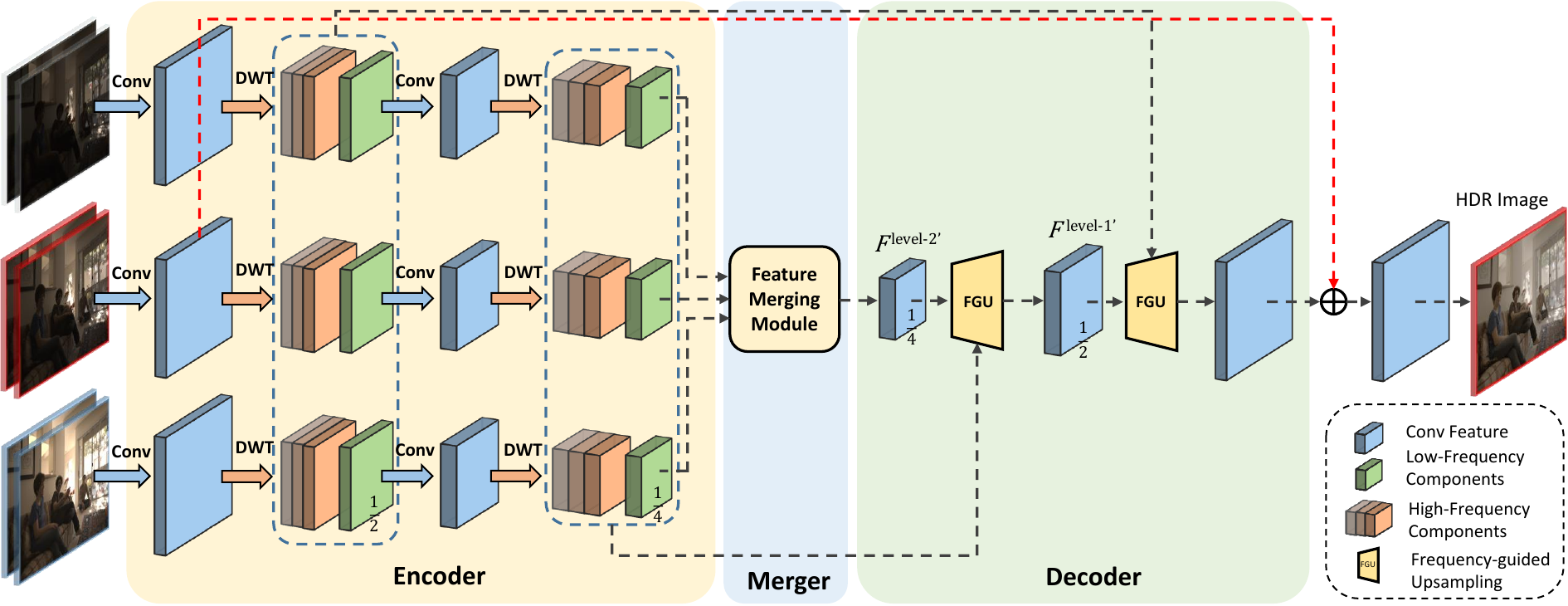}
    \caption{Overall architecture of the proposed FHDRNet. {We consider three pairs of LDR and its HDR as inputs to our FHDRNet and the final reconstructed HDR output is viewed after tone mapping. The proposed network structure contains three parts: an encoder, a merger and a decoder. In the encoder, the input feature maps are decomposed into different frequency sub-bands for further fusion and reconstruction by using DWT. In the merger, the low-frequency sub-bands from the last layer of the encoder are used to generate a single fused feature map. In the decoder, the pre-saved frequency sub-bands are used along with the fused feature map to reconstruct into an upper scale feature map through a frequency-guided upsampling module (FGU). Finally, a global residual connection is used to enhance the feature representation ability of the network. (Best viewed in color.)}}
    \label{fig:overall_structure}
\end{figure*}

In this work, we propose a frequency-guided network (FHDRNet) to explicitly deal with signals of different frequency sub-bands for HDR imaging (see Figure~\ref{fig:overall_structure}). FHDRNet also performs well in the RAW domain. For RAW domain evaluation, we propose a new RAW dataset\footnote{More details about the RAW dataset are included in Section~\ref{sec:dataset}.}. Processing HDR fusion in the RAW domain has the following advantages, especially for wearable devices: 1) From the Image Signal Processing (ISP) pipeline's perspective, it can bring the HDR fusion module to the early stage (\textit{e.g.}, earlier than demosaicing) of the whole ISP pipeline. It can save computations for other modules (\textit{e.g.}, demosaicing) that otherwise have to be done three times, each for one LDR RAW image; 2) RAW data usually have higher bit width (\textit{e.g.},16 bit) and contain more metadata. HDR fusion in the RAW domain will recover more original useful information than that in the RGB domain (8 bit).

The paper's contributions can be summarized as: 
\begin{itemize}
    \item The proposed method - FHDRNet, working in the wavelet domain, is the first to explicitly deal with frequency-specific problems in the HDR imaging task, \textit{e.g.}, ghosting caused by large foreground motion, where the attention mechanism is used on the low-frequency sub-band for fusion to remove such artifacts. The high-frequency sub-bands are used to preserve details (\textit{e.g.}, texture) in the generated HDR image.
    \item A novel frequency-guided upsampling module is proposed to fuse multiple components with different frequency sub-bands from different images into a single set of low and high-frequency sub-bands to upscale the output using Inverse Discrete Wavelet Transform (IDWT).
    \item A new dataset is built for training and evaluating HDR algorithms in the RAW domain, which includes 85 and 15 sets of training and testing samples. This is the first RAW dataset for HDR imaging.
    \item Our method achieves state-of-the-art performance on several public datasets and the new RAW dataset. It also has a good balance between quality and computational efficiency.
\end{itemize} 

\section{Related Work}
In this section, we review the most relevant works, including HDR imaging \citep{khan2006ghost,kalantari2017deep,wu2018deep,yan2020deep,yan2019attention, yan2019multi,yan2021dual,prabhakar2019fast, niu2021hdr} and learning in the wavelet domain \citep{li2020wavelet, abdelhamed2020ntire, liu2018multi}.

\subsection{High Dynamic Range Imaging}
When the scene and camera are completely static, the traditional method \citep{debevec1997} can generate a high quality HDR image by merging them together. But it generates ghosting artifacts when there is motion among the LDR images. Early works \citep{khan2006ghost, pece2010bitmap, li2014selectively} that try to detect and reject the moving pixels fail to handle moving content effectively. To make use of the moving content, ~\citet{bogoni2000extending} and ~\citet{gallo2015locally} first align the input images and then merge the aligned images into one HDR image. 
These methods \citep{bogoni2000extending, gallo2015locally} simply merge the aligned LDR images, and are unable to avoid alignment artifacts. 

Recently, many deep learning-based methods \citep{kalantari2017deep, wu2018deep, yan2020deep} have been developed. ~\citet{kalantari2017deep} propose a deep learning-based model that first aligns the LDR image using optical flow, and then uses a convolutional neural network to generate an HDR image. However, it is difficult to correct the misalignment errors of optical flow, \textit{e.g.}, 
in the moving area, particularly when there also exists occlusion.
%
~\citet{wu2018deep} 
treat the HDR imaging as an image translation problem and use a U-Net to cope with large foreground motion. Though it can reduce the ghosting artifacts, it also blurs image details and hallucinates fine details in the over/under-exposed regions.
%
%
%
~\citet{yan2019multi} adopt three sub-networks with different scales to reconstruct the HDR image gradually. NHDRRNet~\citep{yan2020deep} uses a U-Net to extract features in a low dimension, and then the features are sent into a global non-local network which can fuse the features from inputs according to their correspondence. This method can remove the ghosting artifacts from the final output efficiently. 
AHDR~\citep{yan2019attention} employs an attention mechanism to solve misalignment and avoids the ghosting artifacts. {On the basis of AHDR, DAHDR~\citep{yan2021dual} designs a recurrent spatial and channel attention module to improve the performance.} HDRGAN~\citep{niu2021hdr} proposes a novel adversarial training paradigm to restore missing content in the predicted HDR outputs, utilizing an extra reference-based residual merging block to remove artefacts caused by misalignment. HDRGAN also achieves the state-of-the-art results on the public dataset~\citep{kalantari2017deep}.
SCHDR~\citep{prabhakar2019fast} uses a lightweight optical flow PWC-Net~\citep{sun2018pwc} followed with refinement to align the LDR images first, and then conducts feature aggregation and feature merging to generate an HDR image. 
%
%
These methods fail to explicitly remove the ghosting artifacts and fully exploit the useful information in the inputs.

\subsection{Learning in the Wavelet Domain}
Learning in the wavelet domain has the advantage of explicitly dealing with signals in different frequency sub-bands, and it has been applied to some high-level vision and low-level vision problems, such as classification~\citep{li2020wavelet, williams2018wavelet, ji2012wavelet,cid20173d}, style transfer~\citep{yoo2019photorealistic}, video watermarker~\citep{7563387,9464308}, image denoising \citep{abdelhamed2020ntire,remenyi2014image,ho2012wavelet}, image demoireing \citep{liu2020wavelet}, image deblurring~\citep{7467543}, image/video compression~\citep{suzuki2019wavelet,haghighat2019illumination,9144534}, network compression~\citep{gueguen2018faster}, and super-resolution~\citep{huang2017wavelet, liu2018multi}, etc. 
%
One of the classical image denoising approach is through image shrinkage \citep{donoho1995noising}, where the noisy image is decomposed into low and high-frequency components and then thresholding is applied to the high-frequency coefficients to remove high-frequency noise. For image super-resolution \citep{robinson2010efficient}, the classical approaches are to estimate or interpolate the coefficients of wavelet sub-bands for refining image details.
Recently, DWT has also been applied in deep learning-based image denoising.
The winner of the NTIRE 2020 Denoising Challenge \citep{abdelhamed2020ntire} proposes a multi-level wavelet ResNet for image denoising, where DWT and IDWT are used for downsampling and upsampling.
~\citet{guo2017deep} propose a deep wavelet super-resolution model to recover the residuals of wavelet coefficients of the low resolution image.
~\citet{bae2017beyond} present a wavelet residual network for image denoising and image super-resolution. 
Both \citet{guo2017deep} and \citet{bae2017beyond} only use one level wavelet transformatiom.
~\citet{liu2020wavelet} develop WDNet for image demoireing working directly in the wavelet domain. 
~\citet{liu2018multi} propose a multi-level wavelet-CNN that shows good performance on several image restoration tasks.
%
%

In recent years, discrete wavelet transform (DWT) has also been applied in HDR imaging. ~\citet{omrani2020high} propose a wavelet-based method that aims to use the high-frequency sub-bands obtained from the wavelet decomposition of the input images to recover the details. However, it does not fully utilise the low-frequency sub-band. ~\citet{kaftan2009wavelet} introduce a wavelet-based method to remove noise from the input images with a correlation analysis among them. ~\citet{ramakrishnan2022haar} use Haar wavelet to decompose input images into four different frequency bands. Then, different frequency sub-bands are fused using different predefined fusion rules. Finally, IDWT is applied to achieve the final HDR image using the fused frequency sub-bands. ~\citet{zheng2022domainplus} introduce the cross-transform domain neural network for HDR imaging, which consists of a merging module and a restoration module. In the restoration module, DWT is used to build a cross-transform domain learning block to effectively remove the ghosting artifacts. In the experiments, we observe that in HDR imaging, ghosting artifacts caused by large foreground motion are of low-frequency, while the details are of high-frequency. Since DWT can decompose the input image into different frequency sub-bands, we can take advantage of this property to remove ghosting artifacts or recover details using different frequency sub-bands. Therefore, different from above approaches, our FHDRNet combines wavelet transform with deep learning based method to treat different frequency sub-bands separately. The attention module is used to remove the ghosting artifacts on the low-frequency sub-band and high-frequency sub-bands are used to restore details.

\section{Methodology}
Given a set of LDR images $\{L_{1}, L_{2}, \cdots, L_{n}\}$ with different exposure times, the task of HDR imaging aims to reconstruct an HDR image $H$ that is aligned with the reference frame $L_{\text{ref}}$ (\textit{e.g.}, the medium exposure LDR image). 
In this paper, we follow \citep{kalantari2017deep, wu2018deep, yan2019attention} and use three pairs of LDR and HDR images as input. 
The corresponding HDR images are obtained from the LDR inputs using a gamma correction function:
\begin{equation}
    H_{i} = \frac{L_{i}^{\gamma}}{t_{i}}, i=1,2,3,
\label{eq:gamma_corr}
\end{equation}
where $\gamma$ is set to 2.2 as the default gamma parameter, and $t_{i}$ is the exposure time of $L_{i}$. The final input of the network is the concatenation of the LDR and the corresponding HDR images, forming a 3-pair 6-channel input:
\begin{equation}
    \{I_{1}, I_{2}, I_{3}\} = \{\{L_{1}, H_{1}\}, \{L_{2}, H_{2}\}, \{L_{3}, H_{3}\}\}.
\label{eq:inputs}
\end{equation}

\subsection{Overview of Our Network Structure}
\label{sec:structure}

The proposed network has a U-Net like structure, as shown in Figure \ref{fig:overall_structure}, containing an encoder, a merger and a decoder with skip connections.
In the encoder, the inputs $\{I_{1},I_{2},I_{3}\}$ are sent into three independent sub-networks. In each sub-network, DWT is used for decomposing the feature maps into different frequency sub-bands $\{LL_{i}, LH_{i}, HL_{i}, HH_{i}\}$ ($i=1,2,3$), among which only the low-frequency sub-band $LL_{i}$ is used for the next stage (scale) processing. All frequency sub-bands are also sent to the corresponding frequency-guided upsampling modules through skip connections.
The merger fuses the three inputs (in the low-frequency sub-band) into a ghost-free one, which is then sent to the decoder. 
The network also includes two significant modules: merging module (Section \ref{sec:merging}) and frequency-guided upsampling module (Section \ref{sec:upsampling}). 
The merging module takes only low-frequency components of the previous stage as input and generates a merged result, focusing on structural information. In the decoder, the frequency-guided upsampling module is used to process features in the low-frequency and high-frequency sub-bands independently and then reconstruct the feature maps to a finer scale using IDWT.
A global residual connection is also used to enhance the feature representation ability of the network. The output of the network passes through a tone mapping function (using $\mu$-law) to generate the final tone-mapped HDR image:
\begin{equation}
    \mathcal{T}(H) = \frac{\log(1+\mu H)}{\log(1+\mu)},
\end{equation}
where $H$ is the generated HDR output and $\mu$ is set to 5000 as default to adjust the compression level.


\subsection{Encoder using DWT}
The original inputs $\{I_{1},I_{2},I_{3}\}$ are firstly sent into three independent sub-networks to extract features individually. The features after the first convolution layer (\textit{conv1}) are transformed into different frequency sub-bands through DWT, including one low-frequency component $LL_{i}^{\text{level-1}}$ and three high-frequency components, $\{LH_{i}^{\text{level-1}}, HL_{i}^{\text{level-1}}, HH_{i}^{\text{level-1}}\}$, where $i$ denotes the $i^{th}$ input.
%
According to ~\citet{liu2020wavelet}, the low-frequency sub-band contains more structure information and the high-frequency sub-bands contain more detailed information. 
In order to effectively leverage the decomposed data, the low-frequency component $LL_{i}^{\text{level-1}}$ is used for further decomposition.
In the corresponding frequency-guided upsampling module, the high-frequency components can provide details. So we keep them $\{LH_{i}^{\text{level-1}}, HL_{i}^{\text{level-1}}, HH_{i}^{\text{level-1}}\}$ for reconstruction. 
Then, $LL_{i}^{\text{level-1}}$ goes through the feature extraction (\textit{conv2}) and DWT again. The resulting high-frequency components $\{LH_{i}^{\text{level-2}}, HL_{i}^{\text{level-2}}, HH_{i}^{\text{level-2}}\}$ are kept for later reconstruction, while the low-frequency component $LL_{i}^{\text{level-2}}$ is sent to the feature merging module (Section \ref{sec:merging}) to conduct feature fusion.

\subsection{Merging Module}
\label{sec:merging}

\begin{figure}[t]
    \centering
    \includegraphics[width=1.0\columnwidth]{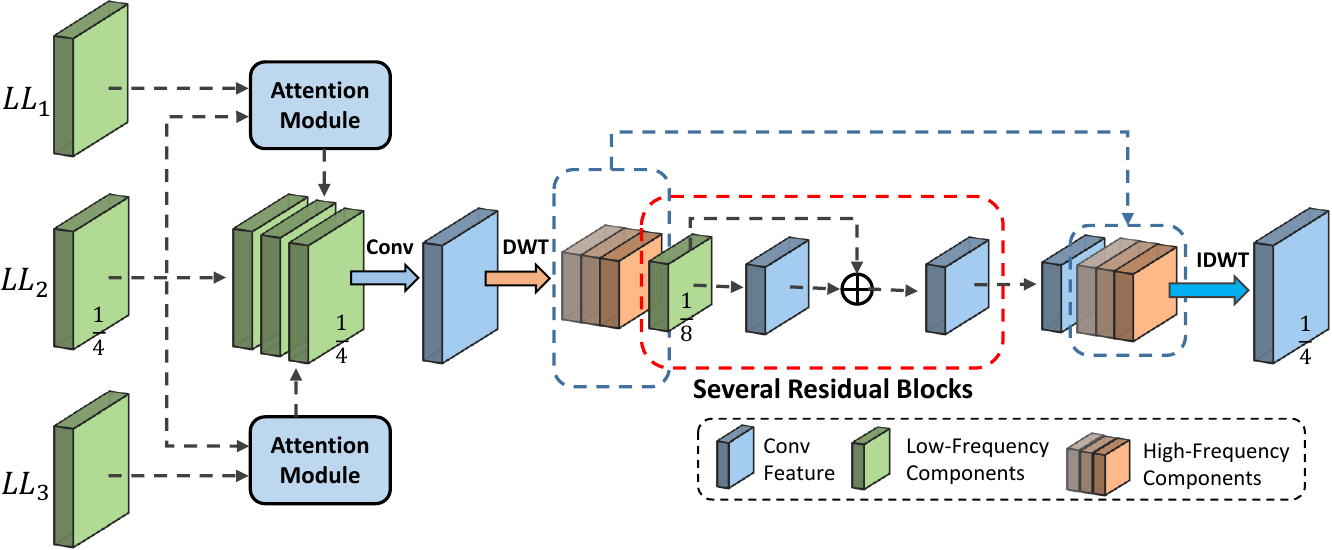}
    \caption{Structure of the merging module. (Best viewed in color.)}
    \label{fig:merge_module}
     \vspace{-5mm}
\end{figure}

The merging module aims at reducing the low-frequency artifacts (\textit{e.g.}, ghosting) by fusing only the low-frequency components (see Figure~\ref{fig:merge_module}).
%
Inspired by AHDR~\citep{yan2019attention}, attention mechanism is applied to deal with the misalignment and saturated regions. 
The support frames $\{LL_{1}^{\text{level-2}}, LL_{3}^{\text{level-2}}\}$ are firstly sent into the attention modules along with the reference frame $LL_{2}^{\text{level-2}}$ to generate corresponding weighted masks $M_{1}$ and $M_{3}$. The attention module includes two convolution layers ($3\times3$ kernel size), with stride and zero padding equal to 1. A sigmoid function is used to normalize the values of the masks to [0, 1].
Next, the feature maps of the support frames are masked and weighted with the masks using element-wise multiplication to get the filtered feature maps $\{LL_{1}^{\text{level-2}^\prime}, LL_{3}^{\text{level-2}^\prime}\}$:
\begin{equation}
    LL_{i}^{\text{level-2}^\prime} = M_{i} \odot  LL_{i}^{\text{level-2}}, i=1, 3,
\end{equation}
where $\odot$ denotes element-wise multiplication. These filtered feature maps and the reference frame's feature maps are concatenated and go through a convolution layer. DWT is applied again to decompose the previous feature map into frequency components with a lower scale for efficient fusion, where the low-frequency component $LL_{i}^{\text{level-3}}$ goes through 9 residual blocks to conduct feature fusion.
Finally, the pre-saved high-frequency components $LH_{i}^{\text{level-3}}, HL_{i}^{\text{level-3}}, HH_{i}^{\text{level-3}}$, and the merged feature are used as the input of  IDWT to recover fused feature maps $F^{\text{level-2}^{\prime}}$ for the frequency-guided upsampling module. 

\begin{figure}[h]
    \centering
    \includegraphics[width=1.0\columnwidth]{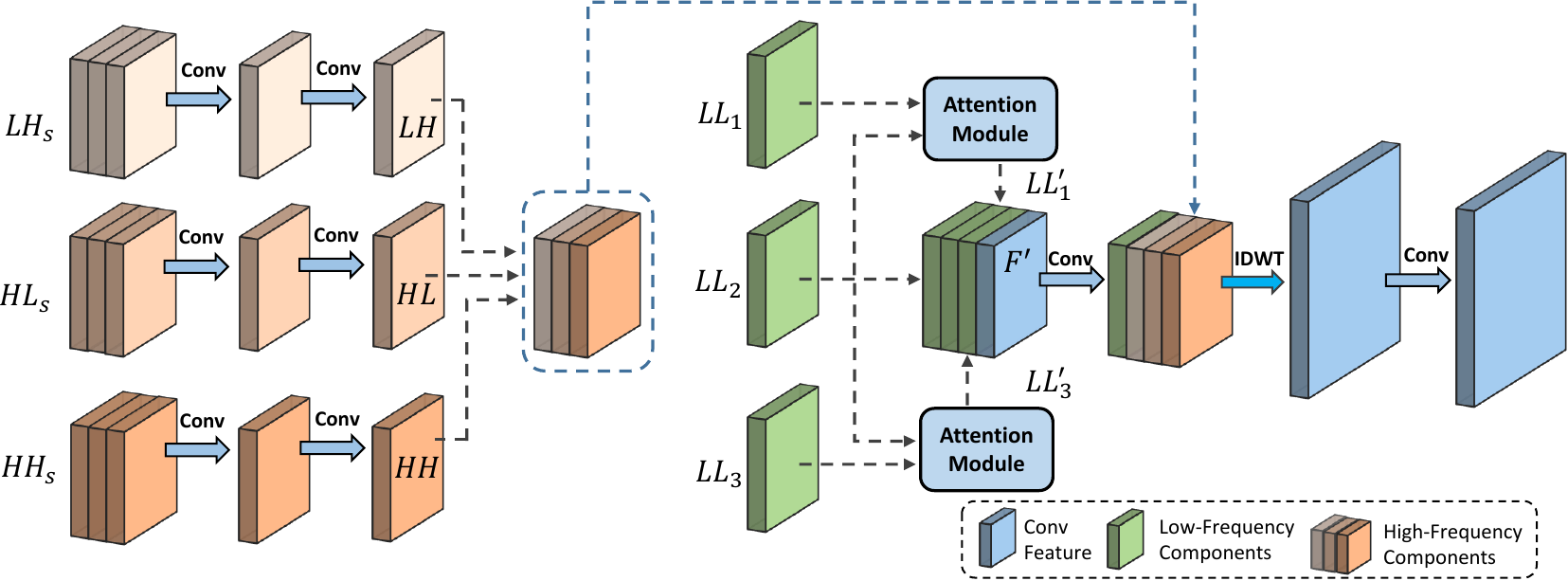}
    \caption{Structure of the frequency-guided upsampling module. (Best viewed in color.)}
    \label{fig:upsample_module}
     \vspace{-5mm}
\end{figure}

\subsection{Frequency-Guided Upsampling Module}
\label{sec:upsampling}
Different from the previous works \citep{liu2018multi, guo2017deep} that use IDWT to reconstruct feature maps from the filtered frequency sub-bands that all go through the same process, our method leverages the decomposed components that go through different processes with the aim of further fusing lower frequency components.
As shown in Figure~\ref{fig:upsample_module}, three sets (each for one input) of decomposed components are used for restoration. Firstly, the high-frequency components are re-grouped into three groups according to their frequency sub-bands: $LH_{s}, HL_{s}, HH_{s}$, where $LH_{s}=\{LH_{1}, LH_{2}, LH_{3}\}$, etc. 
Then, each group is fused by two convolution layers to generate a single set of high-frequency components.
The low-frequency components are fused in a similar way to the merging module by going through the attention modules, 
and $\{LL_{1}^{\prime}, LL_{2}, LL_{3}^{\prime}\}$ along with fused feature maps $F^{\prime}$ (from the previous stage) are concatenated and go though a convolution layer for fusion. 
Finally, IDWT is applied on the fused low and high-frequency components to reconstruct the feature maps. An extra convolution layer is used to squeeze the output's size.

\begin{figure*}[h]
	\centering
	\includegraphics[width=\textwidth]{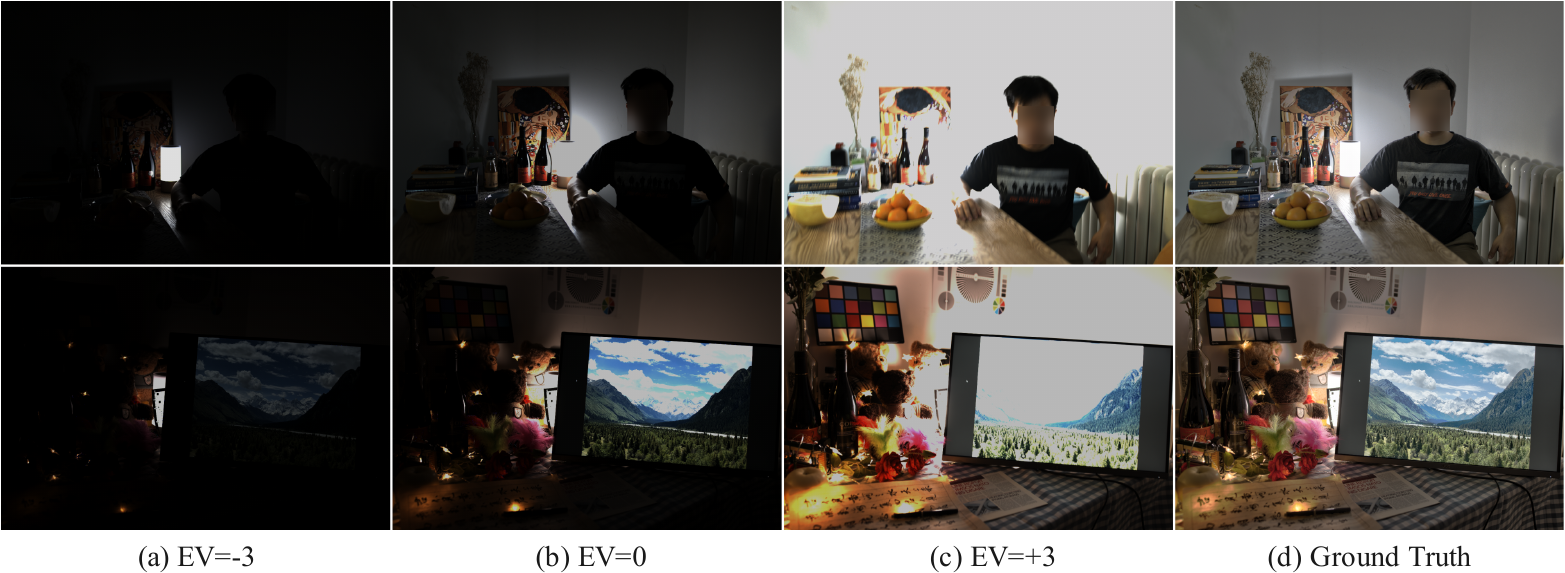}
	\caption{Examples of LDR inputs and corresponding ground truth HDR images in the RAW dataset. (a)-(c) LDR inputs with different exposure bias. (d) Ground truth HDR images. Face has been processed by using Gaussian filter for privacy protection.}
	\label{fig:raw_teaser}
	\vspace{-3mm}
\end{figure*}

\subsection{Training Loss}
Two types of loss are used to train our network: reconstruction loss and Sobel loss. The reconstruction loss is $\ell_{1}$ loss which is the sum of the pixel-wise errors between the generated HDR image and the ground truth. We adopt $\ell_{1}$ loss because it is proved effective for image restoration tasks \citep{yan2019attention}. For the HDR imaging problem, it has been shown that $\ell_{1}$ loss of the tone-mapped images is better than the $\ell_{1}$ loss in the linear space.
The tone mapping function $\mathcal{T(\cdot)}$ is applied to the output to generate the HDR image using $\mu$-law. A basic reconstruction loss is defined as below:
\begin{equation}
    \mathcal{L}_{\mathcal{R}} = \norm{\mathcal{T}(\hat{H}) - \mathcal{T}(H)}_{1},
\end{equation}
where $\hat{H}$ is the predicted HDR linear RGB image and $H$ is the ground truth.

In order to keep the structure information in the generated HDR image, we also use the Sobel loss, which is:
\begin{align}
    \mathcal{L}_{sobel} &= \norm{\nabla_{x}\mathcal{T}(\hat{H}) - \nabla_{x}\mathcal{T}(H)}_{1} \\
    & + \norm{\nabla_{y}\mathcal{T}(\hat{H}) - \nabla_{y}\mathcal{T}(H)}_{1},
\label{eq:sobel_loss}
\end{align}
where $\nabla_{x}$ and $\nabla_{y}$ are the Sobel edge operator in the $x$ direction and $y$ direction respectively. Our final loss is defined as:
\begin{equation}
    \mathcal{L}_{total} = \mathcal{L}_{\mathcal{R}} + \lambda\cdot \mathcal{L}_{sobel},
\end{equation}
and $\lambda$ is a balancing parameter.

\begin{table*}[h]
 \begin{center}
 \caption{Quantitative comparison between the baselines and our proposed network on three public testing datasets: Kalantari~\citep{kalantari2017deep}, Prabhakar~\citep{prabhakar2019fast} and Samsung~\citep{hu2020sensor}. The best and the second best results are \textbf{bold} and \underline{underlined}, respectively. 
 }
 \resizebox{\textwidth}{!}{
    \begin{tabular}{ccccccccccc|c}
      \toprule
        Dataset & Model & Sen & Hu & Kalantari17 & Wu18 &  NHDRRNet & SCHDR & AHDR & DAHDR & HDRGAN &\textbf{Ours}\\
        \midrule
        \multirow{6}{*}{Kalantari} & PSNR-$\mu$ & 40.9453 &32.1872 & 42.7423 &41.6377 &42.4769 & 40.4700 &43.6172 & 43.8400 & \underline{43.8746} & \textbf{43.9066} \\
        &SSIM-$\mu$ &0.9805 & 0.9716 &  0.9877 & 0.9869 &0.9942 & 0.9931 & 0.9956 & 0.9956 & \textbf{0.9958} &\underline{0.9957} \\
        &PSNR-L &38.3147 & 30.8395 & 41.2518& 40.9082 & 40.1978 & 39.6800 & 41.0309 & \underline{41.3100} & 41.0931 &\textbf{41.4736} \\
        &SSIM-L & 0.9749 &0.9511 & 0.9845 &0.9847 & 0.9889 & 0.9899 & 0.9903 & \underline{0.9905} & \textbf{0.9907} &\textbf{0.9907}\\
        &PSNR-PU & 34.4651 & 27.8629 & 36.3597  & 35.8021 & 36.0498 & 36.3154 & 37.2419 & 37.0055 & \underline{37.4420} & \textbf{37.4677}\\
        &SSIM-PU & 0.9783 & 0.9623 & 0.9844 & 0.9811 & 0.9810 & 0.9829 & 0.9848 & 0.9850 & \textbf{0.9862} & \underline{0.9858} \\
        &PSNR-M &30.5507 & 25.5937 & 32.0458 &31.0998 & 34.4113 & 32.3244 & 33.0429 & 33.2900 & \underline{35.2171} &\textbf{35.4163} \\
        &HDR-VDP-2 & 60.5425 & 57.8278 & 64.6319 & 58.3739 & 63.1585 & 62.6192 & \underline{64.8465} & 64.6765 & 64.7617 &\textbf{65.3235}\\
        \midrule
        \multirow{6}{*}{Prabhakar}&PSNR-$\mu$ & 32.7831 & 30.8200 & 35.3400& 31.3100 & 33.0926 & 30.5700 & 33.7200 & \underline{35.3408} & 35.1984 & \textbf{35.5652}\\
        &SSIM-$\mu$ & 0.9740 & 0.9710 & 0.9782 & 0.9733 & 0.9597 & 0.9715 & 0.9789 & 0.9798 & \textbf{0.9829} &\underline{0.9811}\\
        &PSNR-L & 30.4985 & 28.8700 & 32.0800 & 30.7200 & 28.8839 & 31.4400 & 31.8300 & \underline{32.1148} & 30.9183 &\textbf{33.0187}\\
        &SSIM-L & 0.9749 & 0.9564 & \textbf{0.9818} & 0.9518 & 0.9389 & 0.9722 & 0.9674 & \underline{0.9784} & 0.9717 &0.9779\\
        &PSNR-PU & 26.3589 & 23.8109 & \textbf{29.9942} & 26.9620 & 26.3631 & 26.3823 & 27.5110 & 28.8940 & 28.4855 & \underline{29.0431} \\
        &SSIM-PU & 0.9394 & 0.9346 & 0.9477 & 0.9304 & 0.8984 & 0.9356 & 0.9500 & 0.9478 & \textbf{0.9560} & \underline{0.9520} \\
        &PSNR-M & 23.5772 & 27.2642 & 28.4386 & 28.2246 & 27.5843 & 27.7573 & 28.8104 & \underline{29.3938} & 29.3619 &\textbf{29.4924}\\
        &HDR-VDP-2 & 58.4144 & 59.6765 & 62.9073 & 62.4351 & 59.9271 & 62.4376 & 62.3386 & 61.9452 & \underline{62.9463} &\textbf{63.4667}\\
        \midrule
        \multirow{6}{*}{Samsung}&PSNR-$\mu$ & 22.8929 &34.0052 & 23.0547& 41.2544& 41.6741 & 40.1686 & 45.1167 & \underline{45.4359} & 44.0909 & \textbf{45.6199}\\
        &SSIM-$\mu$ & 0.8870 & 0.9896 & 0.8922 & 0.9938 & 0.9942 & 0.9902 & 0.9972 & \textbf{0.9975} & 0.9958 &\underline{0.9974}\\
        &PSNR-L & 24.0611 & 30.3692 & 26.1808 &44.0798 & 43.9048 & 41.4633 & 46.4468 & \underline{47.2869} & 44.9299 & \textbf{48.1514} \\
        &SSIM-L & 0.9541 & 0.9872 & 0.9628 & 0.9976 & 0.9976 & 0.9967 &0.9989 & \underline{0.9990} & 0.9986 &\textbf{0.9991}\\
        &PSNR-PU & 17.4251 & 27.5395 & 17.7897 & 36.5663 & 36.0027 & 34.5844 & 39.3863 & \underline{39.7680} & 38.5820 & \textbf{39.9739} \\
        &SSIM-PU & 0.8545 & 0.9796 & 0.8646 & 0.9883 & 0.9880 & 0.9824 & \underline{0.9943} & \textbf{0.9947} & 0.9926 & \underline{0.9943} \\
        &PSNR-M & 21.2455 & 30.7557 & 16.0222 & 30.9670 & 32.4359 & 28.6001 & 28.8571 & 26.6076 & \underline{32.5280} &\textbf{34.1855}\\
        &HDR-VDP-2 & 55.4090 & 66.2029 & 58.2501 & 71.6399& 70.1484 & 70.7524 & 74.6180 & \textbf{75.0042} & \underline{74.6636} & 74.2816\\
        \bottomrule
    \end{tabular}}
 \label{tab:quantative}
 \end{center}
 \vspace{-5mm}
\end{table*}

\section{Experiments and Results}
\subsection{Datasets}
\label{sec:dataset}
In Section~\ref{sec:introduction}, we have introduced the advantages that processing HDR fusion in the RAW domain. In order to satisfy the requirements that developing HDR imaging algorithms in the wearable devices (\textit{e.g.}, smart phone). We create a new dataset for training and evaluating HDR imaging algorithms in the RAW domain. The data capturing and ground truth merging is according to the method in Kalantari~\citep{kalantari2017deep} and the device is SONY ILCE-7RM2. We capture two sets of images for the same scene: the static set and the dynamic set. Each set contains three images captured with different exposure bias and with high resolution (5120$\times$3456) using RAW format. In the static set, the object is kept static during the capturing, and these images are mainly used to generate the ground truth HDR images. In the dynamic set, the object will do some different movements, and these images are used as inputs for the network. In our dataset, we capture both classic HDR imaging scenes and objective scenes. In order to have an objective evaluation of the generated HDR images, some professional standards are introduced to our dataset, such as Film calibration plate (\textit{e.g.}, details) and SpyderCheckr (\textit{e.g.}, colour). The examples of our dataset are in Figure~\ref{fig:raw_teaser}.

Furthermore, we also provide the corresponding RGB images and metadata (\textit{e.g.}, ISO, F-number,  exposure time, exposure bias, and white balance coefficients) for each set of samples, and these extra data can be used for future works (\textit{e.g.}, training deep learning based end-to-end ISP pipelines). The provided metadata can also be used to calculate the precise exposure ratio (ER) between images, instead of using exposure bias to get an approximate value.

Finally, We capture 253 sets of samples in total and keep 100 sets, where there is no scene motion or object motion in the static sets (\textit{e.g.}, pixel shift is smaller than 5 pixels) for producing better ground truth HDR images. In the experiment, 85 sets of samples are used for training, and 15 sets of samples are used for evaluating. 

The experiments are conducted on four public datasets, including three real datasets (Kalantari dataset~\citep{kalantari2017deep}, Prabhakar dataset~\citep{prabhakar2019fast}), and Tursun dataset~\citep{tursun2016objective}), and one synthetic dataset (Samsung dataset~\citep{hu2020sensor}). Among them, Kalantari, Prabhakar, and Samsung datasets are used for quantitative and qualitative evaluation, while Tursun dataset is used for qualitative comparison only as it does not provide ground truth. In addition, our RAW dataset is also used for training and evaluating the proposed method in the RAW domain.
%
The Kalantari dataset~\citep{kalantari2017deep} includes 74 training samples and 15 testing samples. Each sample contains three LDR images which are captured with different exposure biases: $\{-2, 0, 2\}$ or $\{-3, 0, 3\}$, and the size of each image is 1500$\times$1000. Prabhakar dataset~\citep{prabhakar2019fast} has 116 testing samples
and it is used only for evaluation. The Samsung dataset~\citep{hu2020sensor} is a synthetic one, containing 100 samples. 
The dataset is created in a similar way to the Kalantari dataset, except that all the data are synthesized through a game engine. We choose the first 85 samples for training, and the last 15 for testing. Our RAW dataset is also created in a similar way with higher resolution, and it includes 85 training samples and 15 testing samples.

In the experiments, the training and evaluation are divided into three parts: 1) For real images, the model is trained on the Kalantari~\citep{kalantari2017deep}'s training samples and evaluated on the Kalantari~\citep{kalantari2017deep} and Prabhakar~\citep{prabhakar2019fast} testing samples; 2) For synthetic images, the training and evaluation are on the Samsung dataset~\citep{hu2020sensor}; 3) For RAW images, training and evaluation are on the RAW dataset. For those training samples with ground truth, during training, the images are randomly cropped into 256$\times$256 small patches and then data augmentation (\textit{e.g.}, flip and rotate) is applied for effective training. During evaluation, the entire test images are fed into the network to predict the HDR images. 

\subsection{Experimental Settings}
\textit{1) Implementation Details: } During training, Adam~\citep{kingma2014adam} is selected as the optimizer. The initial learning rate is $2\times10^{-4}$. After 20,000 epochs, it is reduced to $2\times10^{-5}$, and after 20,000 epochs, it is further reduced to $2\times10^{-6}$. We train the network for 60,000 epochs. The batch size is 16. Haar wavelet is used for frequency decomposition. The balancing parameter $\lambda$ is set to 0.25. The code and the dataset are available at:  \href{https://github.com/TianhongDai/wavelet-hdr}{https://github.com/TianhongDai/wavelet-hdr}.

\begin{figure*}[h!]
    \centering
    \includegraphics[width=0.8\textwidth]{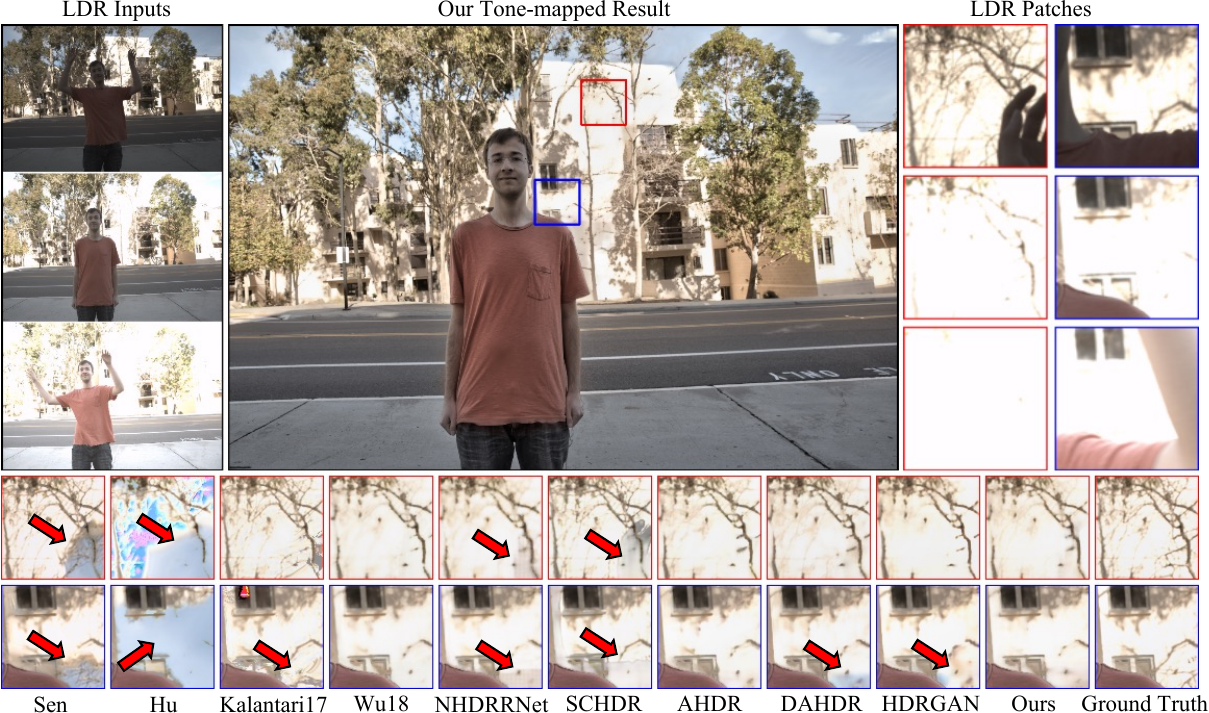}
    \caption{Qualitative comparison between our method and the baselines on the Kalantari testing dataset~\citep{kalantari2017deep}.}
    \label{fig:kalantari_qualitative}
\end{figure*}

\begin{figure*}[h!]
    \centering
    \includegraphics[width=0.8\textwidth]{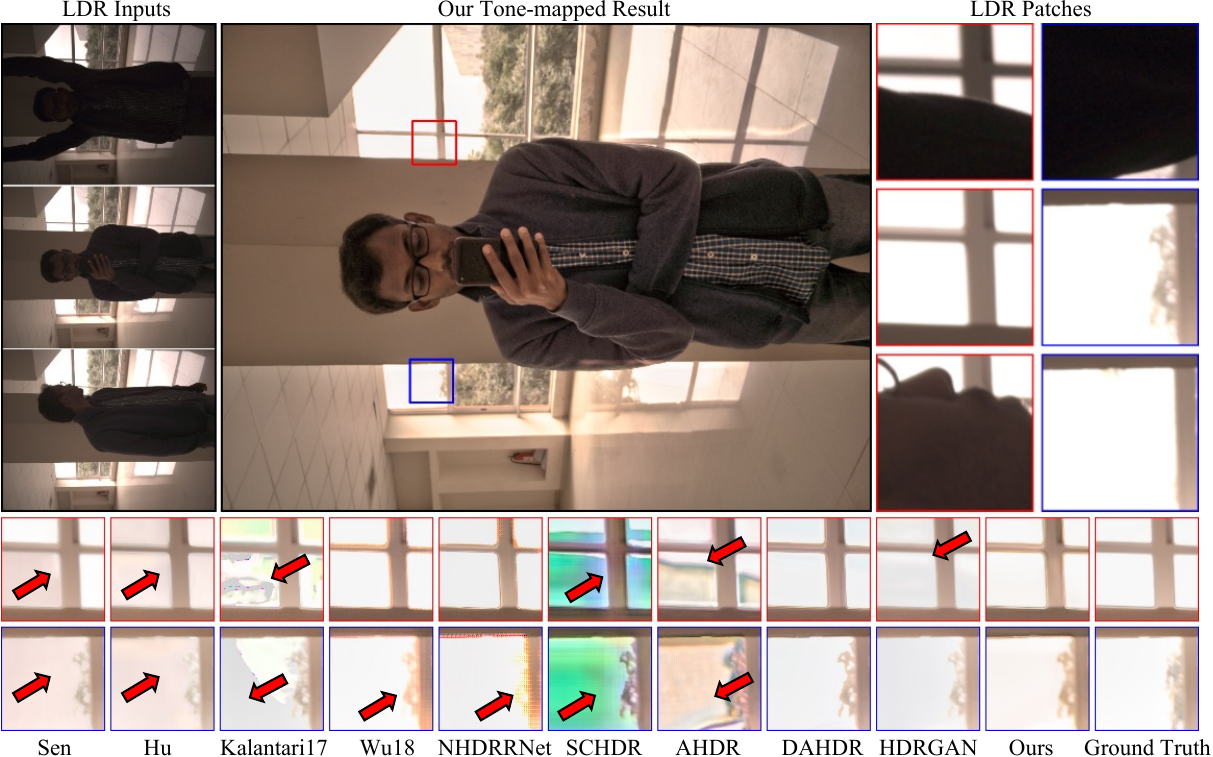}
    \caption{Qualitative comparison between our method and the baselines on the Prabhakar testing dataset~\citep{prabhakar2019fast}.}
    \label{fig:iccp_qualitative}
\end{figure*}

\begin{figure*}[h!]
    \centering
    \includegraphics[width=0.8\textwidth]{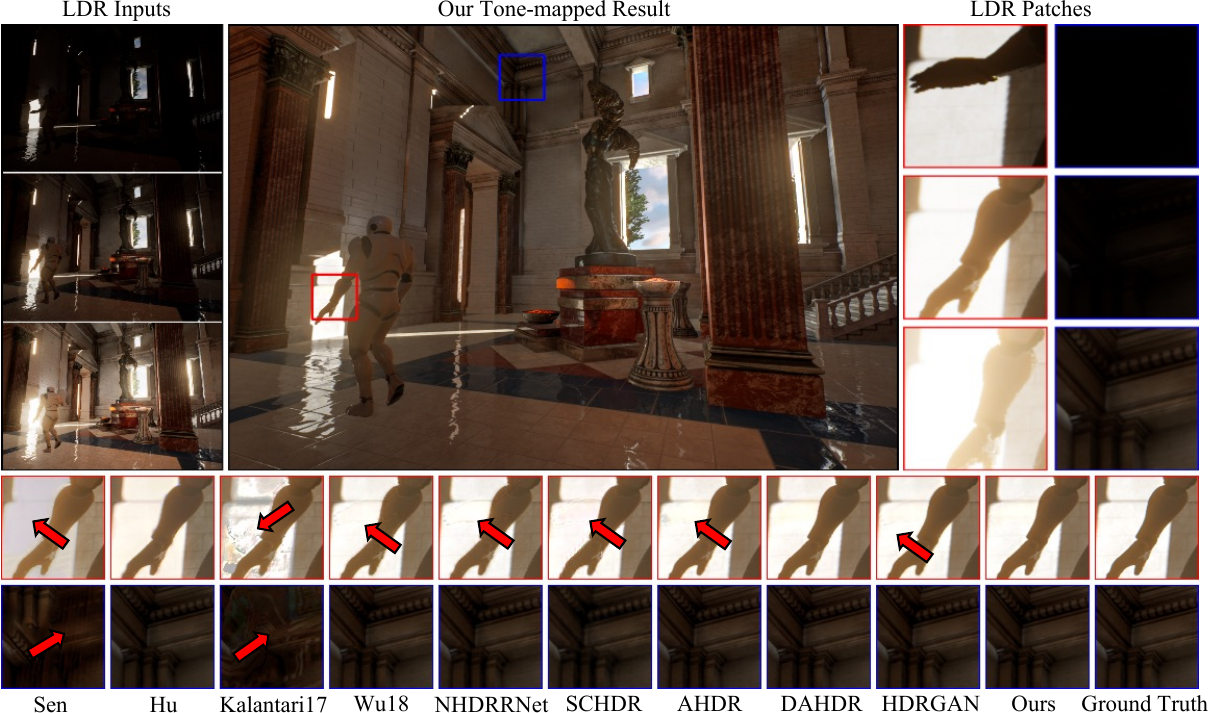}
    \caption{Qualitative comparison between our method and the baselines on the Samsung testing dataset~\citep{hu2020sensor}.}
    \label{fig:samsung_qualitative}
\end{figure*}

\begin{figure*}[h!]
    \centering
    \includegraphics[width=0.8\textwidth]{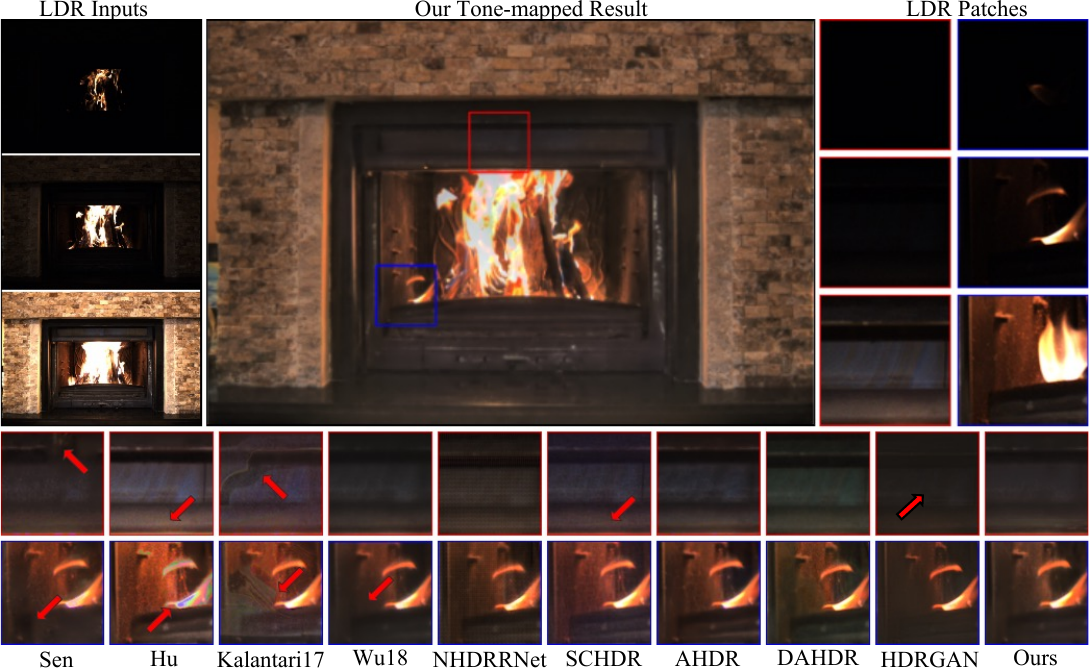}
    \caption{Qualitative comparison between our method and the baselines on the Tursun dataset (Flames)~\citep{tursun2016objective}.}
    \label{fig:tursun1_2}
\end{figure*}
\begin{figure*}[h!]
    \centering
    \includegraphics[width=0.8\textwidth]{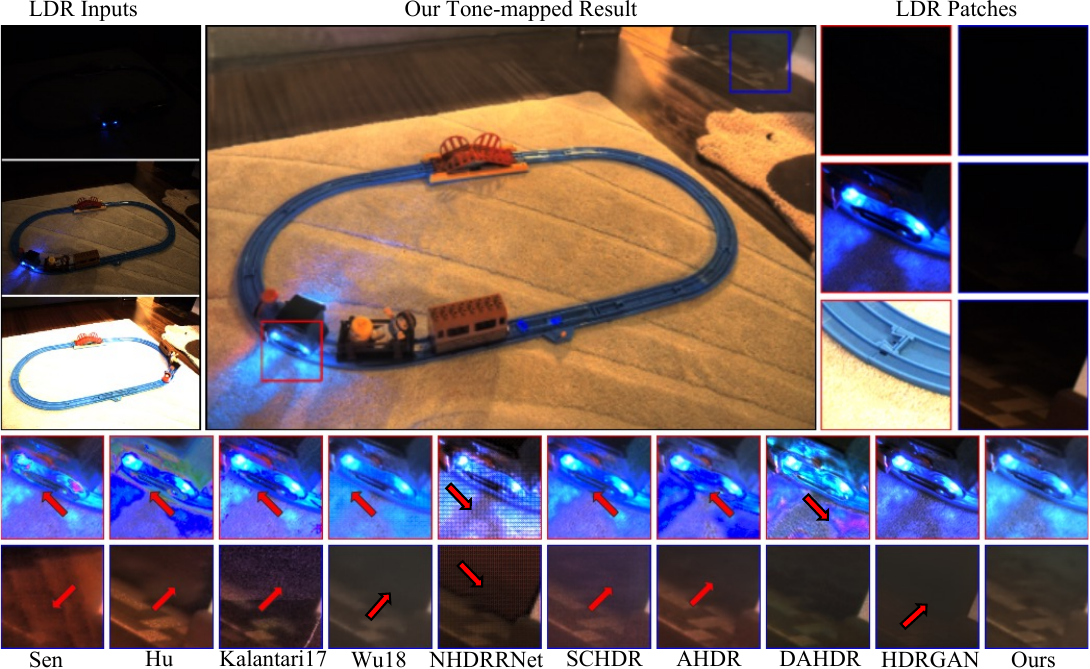}
    \caption{Qualitative comparison between our method and the baselines on the Tursun dataset (ToyTrain)~\citep{tursun2016objective}.}
    \label{fig:tursun1_1}
\end{figure*}

\textit{2) Compared Methods and Evaluation Metrics:} Our model is compared with 9 state-of-the-art methods, including Sen~\citep{sen2012robust}, Hu~\citep{hu2013hdr}, Kalantari17~\citep{kalantari2017deep}, Wu18~\citep{wu2018deep}, NHDRRNet~\citep{yan2020deep}, SCHDR~\citep{prabhakar2019fast}, AHDR~\citep{yan2019attention}, DAHDR~\citep{yan2021dual} and HDRGAN~\citep{niu2021hdr}.
Among them, there are
two patch-based~\citep{sen2012robust,hu2013hdr}, two optical flow based with CNNs~\citep{kalantari2017deep,prabhakar2019fast} and five CNN based without using optical flow~\citep{wu2018deep,yan2019attention,yan2020deep,yan2021dual,niu2021hdr}. For quantitative evaluation, we follow~\citep{kalantari2017deep} to compare PSNR and SSIM results for the linear RGB images and for the tone-mapped images. PSNR-$\mu$ and SSIM-$\mu$ are for the HDR images after tone mapping using $\mu$-law. PSNR-L and SSIM-L are for the HDR images in the linear space. PSNR-PU and SSIM-PU are for the HDR images using the perceptual uniform encoding~\citep{azimi2021pu21}, and the value of peak luminance is 4000. PSNR-M is for the tone-mapped HDR images using the MATLAB built-in function. In addition, we also use HDR-VDP-2~\citep{mantiuk2011hdr}, which is an evaluation metric specially designed to evaluate the visual quality of HDR images, and following parameters are used: display diagonal is 21, viewing distance is 1, and color encoding is ``rgb-native". Furthermore, in order to provide a fair comparison, we re-train HDRGAN~\citep{niu2021hdr} by using the official code with a 256$\times$256 patch size (the original model uses a patch size of 512$\times$512), and other hyperparameters are the same as the default setting.

\begin{figure}[h]
    \centering
    \includegraphics[width=1.0\columnwidth]{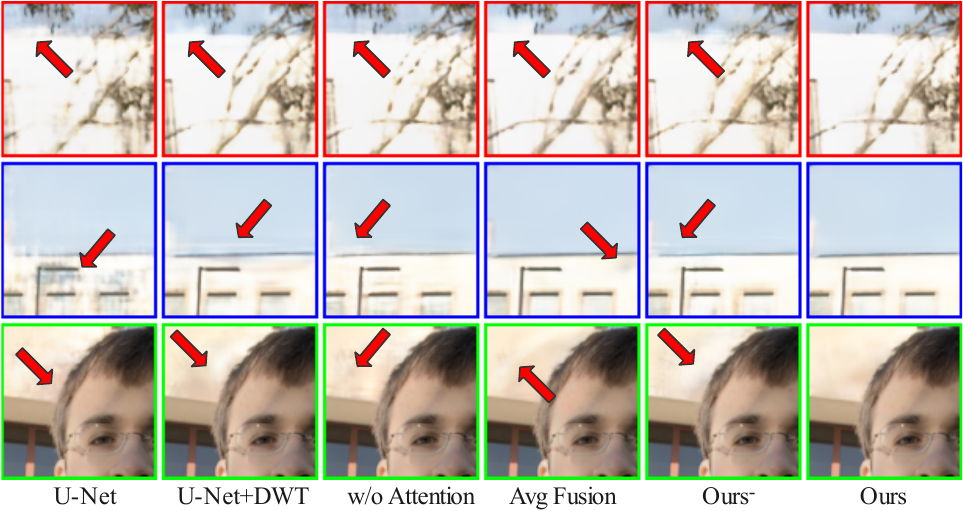}
    \caption{Qualitative comparison between our method and the baseline variants on the Kalantari testing dataset~\citep{kalantari2017deep}.}
    \label{fig:ablation}
    \vspace{-5mm}
\end{figure}

\begin{table*}[h]
 \begin{center}
  \caption{Quantitative results of ablation studies on the Kalantari testing dataset~\citep{kalantari2017deep}. The best and the second best results are \textbf{bold} and \underline{underlined}, respectively.}
  \vspace{-3mm}
 \resizebox{\textwidth}{!}{
    \begin{tabular}{ccc|c|ccc|c|c|c|c}
      \toprule
        Model & U-Net & U-Net+DWT  & w/o Attention & Wavelet:sym2 &  Wavelet:db2 & Wavelet:db3 & Average Fusion & w/o Sobel Loss & $\text{Ours}^{-}$ & \textbf{Ours}\\
        \midrule
        PSNR-$\mu$ & 42.0488 & 42.8238 & 43.4657 & 43.6048 & 43.5820 & 43.4675 & 43.1682 & \underline{43.6257} & 42.8969 & \textbf{43.9066}\\
        SSIM-$\mu$ & 0.9936 & 0.9950 & 0.9953 & \underline{0.9955} & \underline{0.9955} & 0.9954 & 0.9951 & 0.9954 & 0.9947 & \textbf{0.9957}\\
        PSNR-L & 39.0038 & 40.2282 & 40.5313 & 40.8969 & 40.7055 & 40.8254 & 40.7980 & \underline{40.9293} & 40.8455 & \textbf{41.4736}\\
        SSIM-L & 0.9852 & 0.9892 & 0.9895 & 0.9899 & 0.9897 & 0.9901 & \underline{0.9903} & 0.9899 & 0.9899 & \textbf{0.9907}\\
        PSNR-M & 32.6175 & 34.4052 & 34.5538 & 35.3321 & 34.5855 & \textbf{35.7751} & 35.0043 & 35.1833 & 34.8647 & \underline{35.4163}\\
        HDR-VDP-2 & 61.3812 & 63.3845 & 63.5026 & \underline{64.8847} & 64.3280 & 64.4923 & 64.6400 & 64.4789 & 64.5697 & \textbf{65.3235}\\
        \bottomrule
    \end{tabular}}
   \label{tab:ablation}
 \end{center}
 \vspace{-5mm}
\end{table*}

\subsection{Comparison with State-of-the-Arts}
\textit{1) Quantitative Results:} 
Table~\ref{tab:quantative} shows the quantitative comparison between the state-of-the-art models and ours on three datasets: Kalantari~\citep{kalantari2017deep}, Prabhakar~\citep{prabhakar2019fast}, and Samsung~\citep{hu2020sensor}. Our model outperforms most of them, especially on the Kalantari and the Samsung datasets, where it achieves six best and two second best results in eight evaluation metrics on the Kalantari dataset, and achieves five best and two second best results in eight evaluation metrics on the Samsung dataset. For the Prabhakar dataset, our model has four best and three second best results in eight evaluation metrics. HDRGAN~\citep{niu2021hdr} achieves the second best results in most of metrics on the Kalantari dataset, because it uses adversarial learning to restore the missing information in the generated HDR. However, HDRGAN performs poorly on the Samsung dataset, probably due to the sensitivity of the adversarial training to hyperparameters and network structure. {Different from other optical flow free methods using the U-Net structure, AHDR~\citep{yan2019attention} and DAHDR~\citep{yan2021dual} adopt a network structure in a fixed scale. Therefore, AHDR and DAHDR can preserve more information during encoding and merging. With the assistance of the attention mechanism which can detect the misalignment and saturated regions, AHDR and DAHDR achieve the top three scores in most cases. Compared with AHDR and DAHDR, our model consistently outperforms them across all three datasets.}

\textit{2) Qualitative Results:} From Figure~\ref{fig:kalantari_qualitative} to Figure~\ref{fig:tursun1_1}, we show the qualitative comparisons on three public datasets~\citep{kalantari2017deep,prabhakar2019fast,hu2020sensor}. Sen \citep{sen2012robust} and Hu \citep{hu2013hdr} generate strong ghosting artifacts in the images with large foreground motion. 
These traditional methods have worse performances compared with deep learning-based methods.
Optical flow based methods Kalantari17~\citep{kalantari2017deep} and SCHDR~\citep{prabhakar2019fast}, in which the input frames are aligned using optical flow before the further merging operation, benefiting a lot from the explicit alignment. But inaccurate optical flow estimation leads to ghosting artifacts, especially in the areas of large motion (see Figure~\ref{fig:kalantari_qualitative} and Figure~\ref{fig:iccp_qualitative}).
{Wu18~\citep{wu2018deep} and NHDRRNet~\citep{yan2020deep} produce} gridding artifacts (see Figure~\ref{fig:iccp_qualitative} and Figure~\ref{fig:tursun1_1}), because of deconvolution for upsampling. {AHDR~\citep{yan2019attention} and DAHDR~\citep{yan2021dual} also produce ghosts in Figure~\ref{fig:iccp_qualitative} and Figure~\ref{fig:kalantari_qualitative}, respectively.} From these results, our method shows better details than other baselines, because details are preserved in high-frequency sub-bands. Through merging features using low-frequency components with the attention mechanism, the ghosting artifacts are also relieved compared with other methods. 



\begin{table*}[h!]
 \begin{center}
 \caption{Comparison of the running time (second) and GPU memory (GB) with corresponding PSNR-$\mu$ (dB) between the baselines and our method for generating a 1500$\times$1000 HDR image. The sign ``\text{-}" denotes the method is evaluated on a CPU.}
 \resizebox{0.85\textwidth}{!}{
    \begin{tabular}{cccccccccc|c}
      \toprule
        Model & Sen & Hu & Kalantari17 & Wu18 & NHDRRNet & SCHDR & AHDR & DAHDR & HDRGAN &\textbf{Ours}\\
        \midrule
        Time & 51.96 & 293.61 & 68.81 & 0.21 & 0.34 & 0.32 & 0.78 & 0.92 & 0.45 & 0.59\\ 
        Memory & - & - & - & 3.33 & 3.60 & 8.60 & 10.37 & 14.83 & 8.48 & 7.36\\
        PSNR-$\mu$ & 40.95 & 32.19 & 42.74 & 41.64 & 42.48 & 40.47 & 43.62 & 43.84 & 43.87 & 43.91\\
        \bottomrule
    \end{tabular}}
 \label{tab:cost}
 \end{center}
\end{table*}
\begin{figure*}[h]
    \centering
    \includegraphics[width=0.83\textwidth]{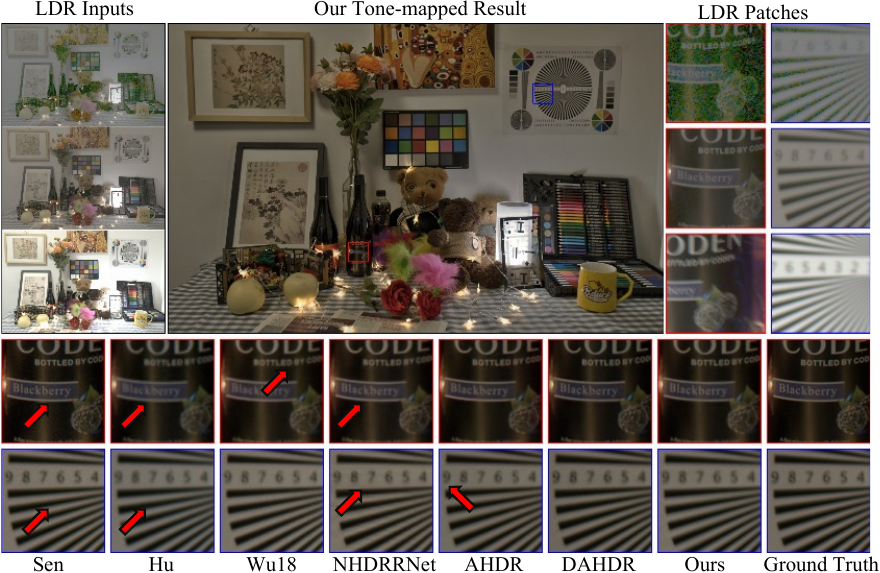}
    \caption{Qualitative comparison between our method and the baselines on the RAW testing dataset. The RAW images are visualized through the same ISP pipeline.}
    \label{fig:raw_test}
\end{figure*}

\begin{table*}[h!]
	\begin{center}
	\caption{Quantitative results between baselines and our proposed network on our RAW dataset. The best and the second best results are \textbf{bold} and \underline{underlined}, respectively.}
		\resizebox{0.65\textwidth}{!}{
			\begin{tabular}{ccccccc|c}
				\toprule
				Model & Sen & Hu & Wu18 & NHDRRNet & AHDR & DAHDR & \textbf{Ours}\\
				\midrule
				PSNR-$\mu$ & 32.0580 & 29.9452 & 37.532 & 37.4309 & 38.7193 & \underline{38.8811} & \textbf{39.1243} \\ 
				SSIM-$\mu$ & 0.9705 & 0.9720 & 0.9869 & 0.9877 & 0.9900 & \underline{0.9903} & \textbf{0.9911} \\
				PSNR-L & 35.7427 & 28.6619 & 39.1129 & 38.6586 & 39.5321 & \underline{39.7531} & \textbf{39.7821} \\
				SSIM-L & 0.9386 & 0.9031 & 0.9944 & 0.9943 & 0.9944 & \underline{0.9947} & \textbf{0.9952} \\
			    PSNR-PU & 25.6776 & 23.1445 & 31.5875 & 31.2615 & \underline{32.4102} & \textbf{32.6657} & 31.9010 \\
			    SSIM-PU & 0.9363 & 0.9366 & 0.9644 & 0.9639 & 0.9712 & \textbf{0.9727} & \underline{0.9716} \\
				PSNR-M & \textbf{29.8978} & 27.9381 & 29.1849 & 28.4560 & 29.0868 & \underline{29.2939} & 29.1621 \\
				HDR-VDP-2 & 65.2848 & 59.7245 & 65.0400 & 65.6121 & 65.6350 & \textbf{65.7841} & \underline{65.6577} \\
				\bottomrule
		\end{tabular}}
		\label{tab:raw}
	\end{center}
	\vspace{-5mm}
\end{table*}

\subsection{Ablation Studies}
In this section, we conduct ablation studies to investigate the contribution of each module in our model on the Kalantari dataset. 
As shown in Table~\ref{tab:ablation}, our ablation studies focus on the following parts: 1) only process low-frequency and high-frequency separately, 2) the importance of the attention mechanism, 3) different types of wavelet, 4) different types of methods to fuse high-frequency components in the upsampling module, 5) the Sobel loss function, and 6) the importance of using ONLY the low-frequency component for further processing (next scale) and merging.

\textit{1) Frequency-Specific Processing:} 
We design a ``U-Net + DWT" model that is basically a U-Net except that it processes the low and high-frequency sub-bands separately. 
The naive replacement leads to an improvement of 0.78dB in terms of PSNR-$\mu$ over the U-Net baseline. This model can outperform several state-of-the-art deep learning models \citep{wu2018deep,yan2020deep,prabhakar2019fast} that adopt U-Net as the backbone.
This is because the high-frequency components can preserve more details. As shown in Figure~\ref{fig:ablation}, the results of ``U-Net + DWT" are smoother and also with better details than U-Net. 

\textit{2) Attention Mechanism:} Inspired by AHDR, attention modules are also used in both the merging module and the upsampling module of the proposed model.
To verify its contribution, we remove the attention modules (indicated as ``w/o Attention" in Table~\ref{tab:ablation}). Compared with our final model, it shows that removing the attention modules {leads to 0.52dB} decrease in terms of PSNR-$\mu$. Different from AHDR and DAHDR which applies the attention to all feature maps of the original scale, we only apply the attention to the low-frequency components of the feature maps on smaller scales (1/8, 1/4, and 1/2). By designing in this way, we specifically align the lower-frequency sub-band to remove ghosting artifacts and also save computation.
As shown in Figure~\ref{fig:ablation}, the results of ``w/o Attention" have ghosting artifacts. 

\textit{3) Types of Wavelet:} In addition to the default Haar wavelet, various types of wavelet are also evaluated: Symlet wavelet (indicated as ``sym2") and Daubechies wavelets with approximation order 2 and 3 (indicated as ``db2" and ``db3"). 
Our model with Haar wavelet outperforms the models with other wavelets. However, using other types of wavelet still gets comparable results, which shows the robustness of our method to the type of wavelet.

\textit{4) Fusion Methods of High-Frequency Components:} Another approach to fuse the high-frequency components in the upsampling module is also investigated. 
Firstly, it groups the components from different inputs with specific frequencies, and then averages the values of these components pixel by pixel to get the fused high-frequency components. 
The average fusion method has worse performance than the CNN fusion. 

\textit{5) Sobel Loss Function:} The Sobel loss contributes 0.28dB improvement for the score of PSNR-$\mu$. It can guide the model to recover better edge information.

\textit{6) Using only Low-Frequency after Decomposition:} To verify our design of using only the low-frequency component after decomposition, all frequency sub-bands are used for the next stage's processing (indicated as ``$\text{Ours}^{-}$"), and it leads to a decrease of PSNR-$\mu$ by 1.01dB. As shown in Figure~\ref{fig:ablation}, the results ($\text{Ours}^{-}$) contain ghosting artifacts. 


\subsection{Trade-off Between Quality and Efficiency}
\label{sec:cost}
High dynamic range (HDR) imaging algorithms are widely used in the real-world devices (\textit{e.g.}, smart phones). Therefore, computational efficiency is also an important factor to evaluate the performance of the algorithms. In this experiment, we test the running time and the memory cost with the corresponding PSNR-$\mu$ scores of the baselines and our method in Table~\ref{tab:cost}. The proposed method needs around 0.59 second to generate an HDR image with 1500$\times$1000 resolution on a RTX-2080Ti GPU, {whereas DAHDR need 0.92 second and has a lower score on PSNR-$\mu$. Besides, DAHDR takes up the most memory among the 8 competitors, because DAHDR merges LDR images in the original scale, and has a ``heavy-weighted" network structure.} 
Furthermore, our method, with 50$\%$ less memory than DAHDR, can still achieve better performance. 
Thus, our method has a good balance between quality and efficiency.



\subsection{Evaluation on the RAW Dataset}
We evaluate FHDRNet and compare with state-of-the-art on our new RAW dataset. As shown in Table~\ref{tab:raw}, {FHDRNet achieves the best performance in terms of PSNR-$\mu$, SSIM-$\mu$, PSNR-L and SSIM-L,} indicating that our model can also be used in the RAW domain. In the qualitative comparison, our method preserves more details in the texture than the baselines (see Figure~\ref{fig:raw_test}), because of the efficient utilization of high-frequency sub-bands. {For example, in Figure~\ref{fig:raw_test}, our method restores better details in the bottle.} Thus, the proposed method can also keep its advantages in the RAW domain.
\subsection{Limitations and Future Work}
Although in this work our proposed FHDRNet outperforms other baselines on the several datasets, and our new created dataset also provides a platform for training and evaluating HDR algorithms in the RAW domain, these two contributions still have some limitations to be addressed in the future.

\textit{1) Other Challenges:} Our proposed FHDRNet and other approaches~\citep{wu2018deep,prabhakar2019fast,yan2019attention,yan2021dual,yan2020deep,niu2021hdr} mentioned in this work mainly focus on mitigating the ghosting artifacts caused by large foreground motion in the multi-frame HDR image reconstruction. However, there are some other types of artifacts that need to be addressed in the multi-frame HDR image reconstruction task~\citep{johnson2015high,mantiuk2015high}:
\begin{itemize}
    \item Blurry Artifacts: It is usually caused by the global camera motion, such as taking photos with a hand-held camera. The instability of the capturing process will lead misaligned images and make the generated HDR image blurry.
    \item Noise Artifacts: This is due to some HDR fusin algorithms that are operated on a per-pixel basis. This means the value of each pixel of the generated HDR image is estimated using the value of the pixel at the same location in in all input images.
    \item Glare Artifacts: It is introduced by the defective camera lens or light streaks around the light source created by special filters mounted on the lens.
\end{itemize}
Thus, in the future, we need to explore more potential solutions so that other types of artifacts can also be addressed at the same time to provide better results.\\

\textit{2) Raw Dataset:} While we capture the images follow a comprehensive pipeline, it still has some aspects to introduce several bias to affect the capability of the models:
\begin{itemize}
    \item Camera Motion: When we capture the LDR images, a tripod is used to stabilise the camera. However, in a real-world scenario, the tripod is not always available to users, which can occasionally lead to large camera motion. In this case, our dataset does not cover this situation effectively and it can affect the final performance of the model. There are some possible solutions to mitigate this problem: 1) the camera can be held in the hand to capture some new data to expand our current dataset. 2) global motion can be added to the existing data manually (\textit{e.g.}, shift images).
    \item Limited Scenarios: In our dataset, the images that captured in the outdoor environment are mostly located in the city center, surrounded by buildings and vehicles. Samples with other outdoor scenarios (\textit{e.g.}, natural scenery) are not included in the current dataset, however capturing photos in various environments is also a requirement for the HDR imaging. To further improve the performance of the model in different scenarios, we will collect more samples in various environments to cover more daily use cases in the dataset.
    \item Limited Size: In this dataset, we keep only 100 high-quality samples for training and evaluation. Although the amount of samples is comparable to the existing datasets, such as Kalantari~\citep{kalantari2017deep}, Prabhakar~\citep{prabhakar2019fast} and Samsung~\citep{hu2020sensor} datasets, it still remains a huge gap compared to the number of samples in datasets from other domains. For example, Flickr2K~\citep{timofte2017ntire} is a dataset for the single image super-resolution task, which consists of 2,650 images with 2K resolution for training and evaluation. The limited training data could affect the generalization ability of the model. Therefore, we will collect more data from different environments to augment our dataset.
\end{itemize}

\section{Conclusion}
In this paper, we have proposed a frequency-guided network (FHDRNet) for high dynamic range (HDR) imaging. In the proposed method, the input LDR images are transformed into the wavelet domain using Discrete Wavelet Transform (DWT). The low-frequency sub-bands are mainly used to avoid ghosting artifacts caused by large motion, while the high-frequency sub-bands are used for preserving details. The attention mechanism is adopted to merge low-frequency information to deal with misalignment. The extensive experiments have shown that our method can remove ghosts and preserve more details. It also achieves state-of-the-art results on several public datasets and our RAW dataset with lower computational costs, compared with previous approaches. We believe it has great potential for more extensive applications of HDR imaging.






\bibliographystyle{model2-names}
\bibliography{refs}



\end{document}